\documentclass
[ reprint,
superscriptaddress,
a4paper,
showpacs,
showkeys
 amsmath,amssymb,
 aps,
prb,
]{revtex4-1}

\usepackage{graphicx}
\usepackage{dcolumn}
\usepackage{bm}

\begin{document}

\preprint{APS/123-QED}

\title{$^{121,123}Sb$ – NQR as a microscopic probe in $Te$ doped correlated semimetal $FeSb_2$  : \\emergence of electronic Griffith phase, magnetism and metallic behavior}

\author{A. A. Gippius}
\affiliation{%
 Max Planck Institute for Chemical Physics of Solids, 01187 Dresden, Germany.
}%
 \affiliation{%
 Faculty of Physics, M.V. Lomonosov Moscow State University, Moscow 119991, Russia.
}%

\author{S. V. Zhurenko}%
\affiliation{%
 Faculty of Physics, M.V. Lomonosov Moscow State University, Moscow 119991, Russia.
}%

\author{R. Hu}
\altaffiliation[Present address: ]{Rutgers Center for Emergent Materials and Department of Physics and Astronomy, Rutgers University, Piscataway, NJ 08854, USA.}
\affiliation{%
 Condensed Matter Physics and Materials Science Department, Brookhaven National Laboratory, Upton, New York 11973, USA.
}%

\author{C. Petrovic}
\affiliation{%
 Condensed Matter Physics and Materials Science Department, Brookhaven National Laboratory, Upton, New York 11973, USA.
}
\author{M. Baenitz}
\email{corresponding author: baenitz@cpfs.mpg.de }
\affiliation{%
 Max Planck Institute for Chemical Physics of Solids, 01187 Dresden, Germany.
}

\date{\today}

\begin{abstract}

$^{121,123}Sb$ nuclear quadrupole resonance (NQR) was applied to $Fe(Sb_{1-x}Te_x)_2$ in the low doping regime (\emph{x = 0, 0.01} and \emph{0.05}) as a microscopic zero field probe to study the evolution of \emph{3d} magnetism and the emergence of metallic behavior. Whereas the NQR spectra itself reflects the degree of local disorder via the width of the individual NQR lines, the spin lattice relaxation rate (SLRR) $1/T_1(T)$ probes the fluctuations at the $Sb$ - site. The fluctuations originate either from conduction electrons or from magnetic moments. In contrast to the semi metal $FeSb_2$ with a clear signature of the charge and spin gap formation in $1/T_1(T)T ( \sim exp/ (\Delta k_BT) ) $, the 1\% $Te$ doped system exhibits almost metallic conductivity and the SLRR nicely confirms that the gap is almost filled. A weak divergence of the SLRR coefficient $1/T_1(T)T \sim T^{-n} \sim T^{-0.2}$ points towards the presence of electronic correlations towards low temperatures. This is supported by the electronic specific heat coefficient $\gamma = (C_{el}/T)$ showing a power law divergence $ \gamma(T) \sim T^{-m} \sim (1/T_1T )^{1/2} \sim  T^{-n/2} \sim C_{el}/T $ which is expected in the renormalized Landau Fermi liquid theory for correlated electrons. In contrast to that the \textit{5\%} $Te$ doped sample exhibits a much larger divergence in the SLRR coefficient showing $1/T_1(T)T \sim T^{-0.72} $. According to the specific heat divergence a power law with $n\ =\ 2\ m\ =\ 0.56$ is expected for the SLRR. This dissimilarity originates from admixed critical magnetic fluctuations in the vicinity of antiferromagnetic long range order with $ 1/T_1(T)T \sim T^{-3/4} $ behaviour. Furthermore $Te$-doped $FeSb_2$ as a disordered paramagnetic metal might be a platform for the electronic Griffith phase scenario.  NQR  evidences a substantial asymmetric broadening of the $^{121,123}Sb$ NQR spectrum for the \emph{5\%} sample. This has purely electronic origin in agreement with the electronic Griffith phase and stems probably from an enhanced $Sb$-$Te$ bond polarization and electronic density shift towards the $Te$ atom inside $Sb$-$Te$ dumbbell. 

\end{abstract}

\maketitle


\section{Introduction}
Magnetic resonance is a very suitable microscopic tool for correlated matter at the verge of long range magnetic ordering and aims in particular to expose the real nature of the magnetic fluctuations (antiferromagnetic (\textit{afm}) -  versus ferromagnetic (\textit{fm})) by temperature- and field- scaling \cite{9780198517672}. Local moment \textit{4f}- and \textit{5f}- systems driven by RKKY - and Kondo – interaction could be tuned towards order through the quantum critical point (QCP) by either pressure, substitution or magnetic field \cite{Doniach1977,Stewart_2001,Gegenwart_2008,L_hneysen_2007,Si_2010,9780521514682,Brando_2016,Miranda_2005}. Among \textit{3d}- magnets tunable quantum criticality could be found in itinerant systems like $NbFe_2$ \cite{Brando_2008} and $(Ta,V)Fe_2$ \cite{Horie_2010, Brando_2016_1} but also in systems with more localized $Fe$ moments like $YFe_2Al_{10}$ \cite{Khuntia_2012} and $YbFe_2Al_{10}$ \cite{Khuntia_2014}. Here in contrast to the itinerant $Fe$ systems there is strong evidence for the emergence of weak Kondo interaction among the localized $Fe$ moments. Signatures of “Kondo – type” of correlations are also found in some magnetic semimetals. $FeSi$ \cite{Jaccarino_1967,Mandrus_1995,Manyala_2004}, $FeSb_2$ \cite{Petrovic_2005,Gippius_2014} and $FeGa_3$ \cite{Umeo_2012, Gippius_2014_PRB, Majumder_2016} attracted great attention because of their non-magnetic ground state and their promising low temperature thermoelectric performance. Metallic behavior and $Fe-$based magnetism could be introduced by controlled substitutions on the $Fe-$ or the framework- site. For example for $Fe(Ga_{1-x}Ge_x)_3$ $Ga-$NQR was performed to monitor the effect of $Ge-$ substitution across the phase diagram and to probe the magnetic fluctuations at zero magnetic field via the spin lattice relaxation rate (SLRR) trough the QCP. In conclusion we found an absence of induced disorder, localized antiferromagnetic Kondo-like correlations at low doping levels and critical ferromagnetic fluctuations at the QCP \cite{Majumder_2016}. In contrast to that the $Co$ substitution on the $Fe$ site introduces antiferromagnetic correlations in $(Fe_{1-x}Co_x)Ga_3$ but with sizable induced disorder \cite{Gippius_2014_PRB, Verchenko_2012}. Along this line we started to work on $Sb$ NQR in $Fe(Sb_{1-x}Te_x)_2$ where in contrast to $Fe(Ga_{1-x}Ge_x)_3$ an electronic Griffith phase is predicted for the disordered paramagnetic metal at the verge of canted antiferromagnetism  \cite{Hu_2009, Hu_2012}. Being a local probe at zero field NQR could capture both most relevant points a) the degee of disorder b) the onset of critical antiferromagnetic fluctuations at the verge of long range order. Furthermore NQR might be of use to disentangle the electronic from the magnetic Griffith phase \cite{Miranda_2005}. In the correlated electron metal picture the SLRR is strongly related to the specific heat coefficient $\gamma$  via $ 1/T_1T \sim N^2(E_F) \sim \gamma ^2 \equiv (C/T)^2 $  (Korringa law \cite{9780198517672}). For weak itinerant metals the Moryia- and the Herz Millis theory captures many different cases and systems \cite{9783642825019, Hertz_1976, Millis_1993}. Especially at the verge to long range magnetic order power laws for the SLRR are predicted $(1/T_1(T)T \sim T^{-3/4}$ (\textit{afm}) $\sim T^{-4/3} $ (\textit{fm}) \cite{9783642825019, Hertz_1976, Millis_1993} ). Frequently non Fermi liquid behavior (NFL) was found for many systems which frequently originates from by local disorder. Disorder induced NFL behavior is discussed for many \textit{3d}- and \textit{4f}- and \textit{5f}- system \cite{Miranda_2005}. Here for itinerant \textit{3d} systems (and to some extend  $U-$ based \textit{5f} systems) the Griffith phase was established \cite{Miranda_2005, Castro_Neto_1998, Griffiths_1969, Vojta_2010, Ubaid_Kassis_2010} whereas for some more localized \textit{4f} Kondo systems the Kondo glass scenario was proposed to capture the effect of disorder on the bulk properties \cite{Westerkamp_2009}. So far detailed microscopic studies like NMR or $\mu$SR in quantum critical itinerant \textit{d} electron semimetals on the local effect of doping and the evolution of an electronic Griffith phase are missing. Here we report the results of $^{121,123}Sb$ NQR spectroscopy and nuclear spin-lattice relaxation (SLR)$\ 1/T_1$ experiments on the correlated semimetal $FeSb_2$ and the $Te$ doped systems $Fe(Sb_{0.99}Te_{0.01})_2$ and $Fe(Sb_{0.95}Te_{0.05})_2$.

\section{Experimental}
Single crystals of $Fe(Sb_{1-x}Te_x)_2$ (\textit{x = 0.01, 0.05}) were prepared as described in \cite{Hu_2009}. For NQR measurements $Fe(Sb_{1-x}Te_x)_2$ single crystals which exhibit good metallic conductivity already at \textit{x = 0.01} \cite{Hu_2009} were crushed into fine powder and mixed with paraffin. NQR experiments were performed using phase-coherent pulsed Tecmag-Apollo NMR spectrometer. $^{121,123}Sb$ NQR spectra were measured using a frequency step point-by-point spin-echo technique at 4.2 K by integration of the spin-echo envelope in the time domain and averaging over scan accumulation number which depends on the sample. The $^{123}Sb$ nuclear spin-lattice relaxation was measured using the saturation recovery method in the temperature range of 2.5 - 200 K. In addition, low temperature specific heat measurements were carried out on $Fe(Sb_{1-x}Te_x)_2$ (\textit{x = 0.01, 0.05}) single crystals using the Quantum Design PPMS in the temperature range of 0.5 - 30 K.

\section{Results}

\subsection{$^{121,123}Sb$ NQR spectra}
Bulk measurements reported previously \cite{Hu_2009, Hu_2012} provide only macroscopic evidence for the emergence of an electronic Griffith phase accompanied by NFL behavior in $Te$-doped $FeSb_2$. To obtain a microscopic insight into underlying physics of this system we performed $^{121,123}Sb$ nuclear quadrupole resonance (NQR) spectroscopy study on the same $Fe(Sb_{1-x}Te_x)_2$ (\textit{x = 0.01, 0.05}) samples. $^{121,123}Sb$ NQR spectra measured at 4.2 K for both samples are presented in Fig.1 together with the spectrum of the undoped $FeSb_2$ at 10 K adopted from \cite{Gippius_2014}. As seen from this figure even a very small \textit{(1\%)} $Te$ doping causes significant broadening of the $Sb$ NQR lines. Moreover, $^{121}Sb\ \nu_1$ line (58.5 MHz; $\mid\pm1/2\rangle\leftrightarrow\mid\pm3/2\rangle$ transition) and $^{123}Sb\ \nu_2$ line (55.8 MHz; $\mid\pm1/2\rangle\leftrightarrow \mid\pm3/2\rangle$ transition) already start to overlap in the $Fe(Sb_{1-x}Te_x)_2$ ($x\ =\ 0.01$) compound. Further increase of $Te$ doping (\textit{x=0.05}) leads to complete overlapping of these two NQR lines and formation of two broad shoulders to the left from $^{123}Sb\ \nu_2$ NQR line. Similar asymmetric broadening with formation of a low frequency shoulder is exhibited by all other $^{121,123}Sb$ NQR transition lines in the $Fe(Sb_{0.95}Te_{0.05})_2$ sample (see Fig.1, upper panel).
\begin{figure}[h]
\includegraphics[width=\linewidth]{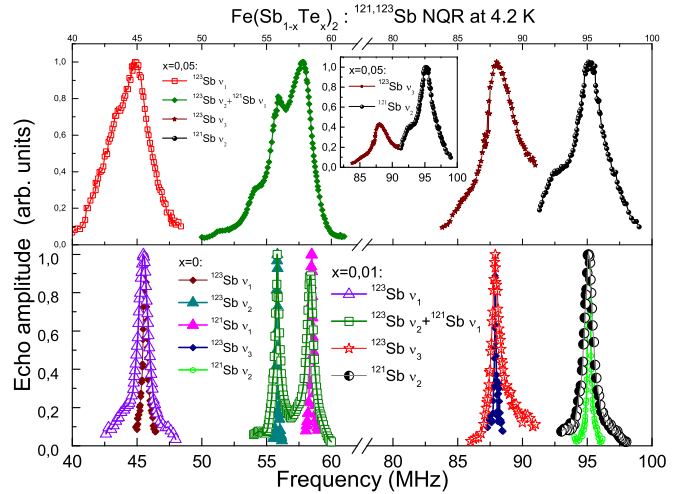} 
\caption{\label{fig:epsart} (color online). $^{121,123}Sb$ spectra measured at 4.2 K in $Fe(Sb_{1-x}Te_x)_2$ compounds with \textit{x = 0.01}(lower panel) and \textit{0.05} (upper panel). For comparison, the same $Sb$ NQR lines for the undoped $FeSb_2$ measured at 10 K and retrieved from \cite{Gippius_2014} are presented (lower panel). The intensities of all transitions except $\nu_2$ line (55.9 MHz; $\mid\pm3/2\rangle\leftrightarrow \mid\pm5/2\rangle$ transition) for the $Fe(Sb_{0.95}Te_{0.05})_2$ sample are normalized on their maximum intensity values. Inset: $^{123}Sb$ $\nu_3$ line (88.0 MHz; $\mid\pm5/2\rangle\leftrightarrow \mid\pm7/2\rangle$ transition) and $^{122}Sb$ $\nu_2$ line (95.1 MHz; $\mid\pm5/2\rangle\leftrightarrow \mid\pm7/2\rangle$ transition) without normalization for the $Fe(Sb_{0.95}Te_{0.05})_2$ sample. Solid lines are guides for eye.}
\end{figure}
    \newline \indent The full width at half maximum (FWHM) for $^{123}Sb$ $\nu_1$ line (44.85 MHz; $\mid\pm1/2\rangle\leftrightarrow\mid\pm3/2\rangle$ transition) amounts 0.45 MHz, 0.91 MHz and 3.63 MHz for the $Te$ concentration \textit{x = 0, 0.01} and \textit{0.05}, respectively. In other words, \textit{only 5\%} of heterovalent doping of $Te$ for $Sb$ results in almost one order of magnitude $Sb$ NQR line broadening which is rather substantial. For comparison, \textit{5\%} of $Co$ substitution for $Fe$ in relative to $FeSb_2$ nonmagnetic Kondo-like semiconductor $FeGa_3$ causes increasing of $^{69}Ga$ (I=3/2) NQR FWHM from 0.044 MHz to 0.18 MHz \cite{Likhanov_2016} which is factor of 2 less than that in $FeSb_2$. Unfortunately, we were not able to estimate FWHM values for other $^{121,123}Sb$ NQR lines due to line overlapping in the $Fe(Sb_{0.95}Te_{0.05})_2$ sample. In order to extract quantitative information from experimental $^{121,123}Sb$ NQR spectra we determined the line width at 80\% level from maximum line intensity. The obtained values are listed in Table 1 demonstrating considerable increase in $^{121,123}Sb$ NQR line width in $FeSb_2$ with $Te$ doping.

\begin{table*}
\caption{\label{tab:table 1} Width of the $^{121,123}Sb$ NQR transition lines in $Fe(Sb_{1-x}Te_x)_2$ samples determined at 80\% from maximum line intensity.}
\begin{ruledtabular}
\begin{tabular}{c|c|c|c|c|c}
 &\multicolumn{2}{c|}{$^{121}Sb$}&\multicolumn{3}{c}{$^{123}Sb$}\\
 &\multicolumn{2}{c|}{$I=5/2$}&\multicolumn{3}{c}{$I=7/2$}\\
 &\multicolumn{2}{c|}{$\gamma / 2 \pi = 10.188$ MHz/T}&\multicolumn{3}{c}{$\gamma / 2 \pi = 5.517$ MHz/T}\\
 &\multicolumn{2}{c|}{$Q=-0.36$ Barn}&\multicolumn{3}{c}{$Q=-0.49$ Barn}\\
 $Fe(Sb_{1-x}Te_x)_2$&$\Delta \nu _1$,MHz &$\Delta \nu _2$,MHz  &$\Delta \nu _1$,MHz &$\Delta \nu _2$,MHz &$\Delta \nu _3$,MHz \\ \hline

 $x=0$&$0.11$&$0.14$&$0.19$&$0.07$&$0.06$ \\ \hline
 $x=0.01$&$0.34$&$0.38$&$0.49$&$0.13$&$0.30$\\ \hline
 $x=0.05$&$1.39$&$1.17$&$1.53$&$-$&$1.48$\\

\end{tabular}
\end{ruledtabular}
\end{table*}

\subsection{$^{123}Sb$ nuclear spin-lattice relaxation}
To probe the effect of small $Te$ doping on the dynamical properties of $FeSb_2$ system we performed $^{123}Sb$ nuclear spin-lattice relaxation (SLR) measurements at $^{123}Sb$ $\nu_2$ NQR line ($\mid\pm3/2\rangle\leftrightarrow\mid\pm5/2\rangle$ transition) as a function of temperature in the range of 2.5 - 200 K by means of saturation recovery method. We have selected this line to enable comparison with the SLR data for the undoped $FeSb_2$ semiconductor available only for $^{123}Sb$ $\nu_2$ NQR line \cite{Gippius_2014}. Since only one NQR transition line was saturated, the $^{123}Sb$ (I = 7/2) magnetization recovery curves for $Fe(Sb_{1-x}Te_x)_2$ (\textit{x = 0.01}, \textit{0.05}) samples were fitted by the sum of three stretched exponents \cite{Gippius_2014, Chepin_1991}:
\begin{equation}
 M(\tau)=M_0+\sum\limits_{i=1}^3 C_i[1-exp(-(2k_iW_i\tau)^n)]
\end{equation}
Here M($\tau$) is the spin-echo integrated intensity, $M_0$ is the remaining magnetization after the saturation comb (at $\tau\rightarrow$\ 0),$\ \tau$ is the delay time between the saturation comb and the spin-echo pulse sequence, $2W_0 = 1/T_1$ is the $^{123}Sb$ nuclear spin-lattice relaxation rate, $C_i(\eta)$, $k_i(\eta)$ are the weighting coefficients. The values of $C_i(\eta)$, $k_i(\eta)$ for $FeSb_2$ ($\eta$ = $0.43$) were taken from the numerical calculations \cite{Chepin_1991} and were assumed not affected by $Te$ doping. The stretched exponent parameter $n$ was introduced in Eq. (1) to account for the structural disorder caused by $Te$ doping. The examples of experimental recovery curves and their best fits to equation (1) obtained for $Fe(Sb_{1-x}Te_x)_2$ (\textit{x = 0.01, 0.05}) samples at 4.2 K compared with that for the undoped $FeSb_2$ at 10 K (retrieved from \cite{Gippius_2014}) are presented in Fig.2. As seen from this Figure, the approximation of the experimental recovery curves to equation (1) is rather good. While for binary $FeSb_2$ $n \equiv 1$, increasing of $Te$ doping leads to significant decrease of the stretched exponent parameter: \textit{n = 0.73(1)} for \textit{x = 0.01} and \textit{n = 0.64(2)} for \textit{x = 0.05}. This effect is a consequence of spatial distribution of $1/T_1$ values due to growing structural and magnetic disorder in $FeSb_2$ crystal lattice caused by $Te$ substitution.
    \newline \indent It is worth to mention two characteristic features seen from Fig.2. First, the remaining magnetization $M_0$ after the saturation comb (at $\tau \rightarrow$ 0) is dramatically increasing with $Te$ doping \textit{x}: while the initial saturation is almost perfect in the undoped $FeSb_2$ ($M_0 \approx 0.04$), $M_0$ becomes $\approx$ $0.3$ for \textit{x = 0.01} and accomplishes $\approx\ 0.54$ for \textit{x = 0.05}. This effect reflects an extreme broadening and even overlapping of $Sb$ NQR lines in $Fe(Sb_{1-x}Te_x)_2$ with increasing \textit{x} (Fig.1). The intense spin diffusion effectively hampers saturation process despite all our efforts to optimize the saturation comb and minimize the $M_0$ value. The second interesting feature of the experimental data shown in Fig.2 is significant visible shift of the recovery curves towards low $\tau$ values with increasing $Te$ doping \textit{x} which indicates very fast increase of the $1/T_1$ values with increasing {x}. This effect also favors increasing of the remaining magnetization $M_0$, as have been observed for the undoped $FeSb_2$ sample with increasing temperature \cite{Gippius_2014}.
\begin{figure}[h]
\includegraphics[width=\linewidth]{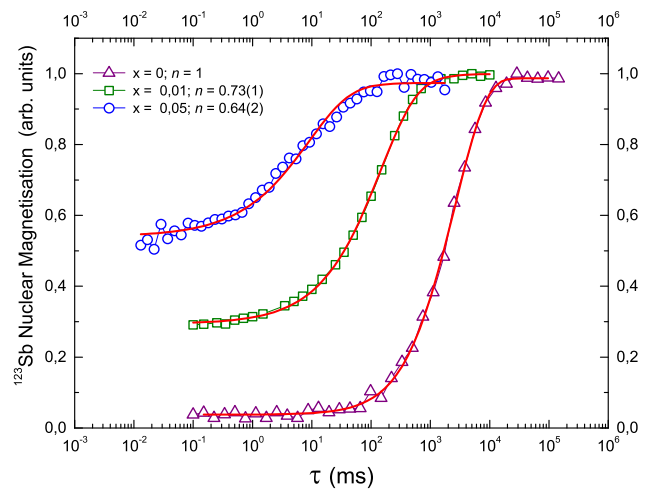} 
\caption{\label{fig:epsart} (color online). $^{123}Sb$ magnetization recovery curves for the $\nu_2$ NQR line (quadrupole transition $\mid\pm3/2\rangle\leftrightarrow \mid\pm5/2\rangle$) in the $Fe(Sb_{1-x}Te_x)_2$ (\textit{x = 0.01, 0.05}) samples at 4.2 K and $FeSb_2$ at 10 K. The latter curve was adopted from Ref.\cite{Gippius_2014}. Solid lines are the best fits to Eq.(2) with \textit{n = 1, 0.73(1), 0.64(2)} for \textit{x = 0, 0.01, 0.05}, respectively.}
\end{figure}
\newline \indent The resulting temperature dependences of $1/T_1T$ as a function of temperature for $Fe(Sb_{1-x}Te_x)_2$ (\textit{x = 0, 0.01, 0.05}) samples are presented in Fig.3. As clearly seen from this figure, even low (\textit{x = 0.01}) $Te$ doping leads to drastic increase of the $Sb$ NSLR in more than one order of magnitude in the low temperature range 2 - 50 K. As has been shown in Refs.\cite{Hu_2009, Hu_2012}, even extremely low $Te$ doping of \textit{x = 0.001} leads to transition from semiconducting to metallic behavior so that at \textit{x = 0.01} one can expect Korringa-like SLRN governed by conduction electrons. Indeed, for the $FeSb_{0.99}Te_{0.01}$ sample $1/T_1T(T)$ might be considered as almost temperature independent in the range of 2 - 70 K (Fig.3). Above 70 K $1/T_1T$ in $FeSb_{0.99}Te_{0.01}$ sample increases merging to that for the undoped $FeSb_2$.
\begin{figure}[h]
\includegraphics[width=\linewidth]{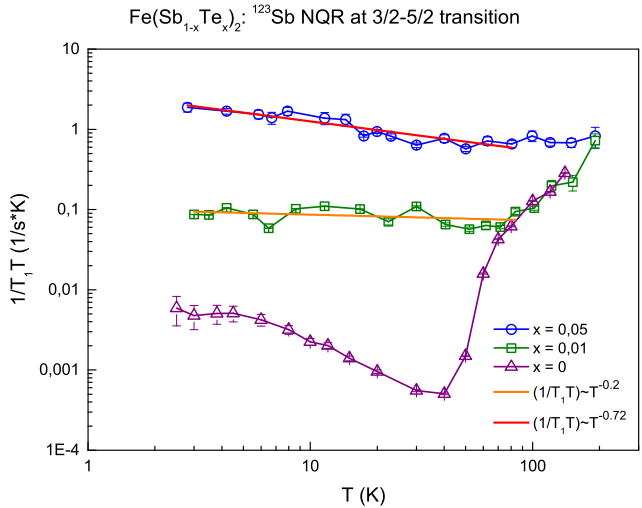} 
\caption{\label{fig:epsart} (color online). $1/T_1T$ as a function of temperature for the $^{123}Sb$ $\nu_2$ NQR line ( $\mid\pm3/2\rangle\leftrightarrow \mid\pm5/2\rangle$ ) in $Fe(Sb_{1-x}Te_x)_2$ compounds (\textit{x = 0, 0.01} and \textit{0.05}). Solid straight lines are the best linear fits according to formula: $1/T_1T = a*T^{-2(1+\lambda)}$ (see text).}
\end{figure}
    \newline \indent For the $FeSb_{0.95}Te_{0.05}$ sample where $1/T_1T$ is one order of magnitude higher than for $FeSb_{0.99}Te_{0.01}$ and a power-law divergence $1/T_1T \sim T^{-0.72}$ ($1/T_1 \sim T^{0.28}$) (Fig.3) was found towards low temperatures.
\begin{figure}[h]
\includegraphics[width=\linewidth]{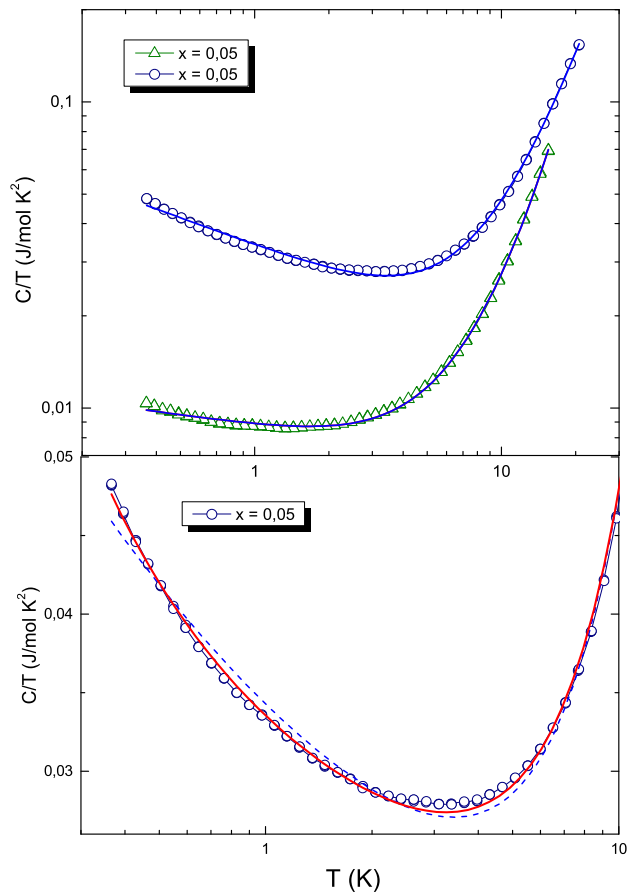} 
\caption{\label{fig:epsart} (color online). Upper panel: \textit{C/T} vs. \textit{T} plot in $Fe(Sb_{1-x}Te_x)_2$ compounds (\textit{x = 0.01} and \textit{x = 0.05}). Solid lines are the best fits to Eq. (1) (see text).
Lower panel: Low temperature part of the \textit{C/T} vs. \textit{T} plot for the $Fe(Sb_{0.95}Te_{0.05})_2$ sample. Dashed and solid lines are the best fits to Eqn.(2) and (5), respectively.}
\end{figure}

\subsection{Specific heat}
In addition to the NQR spectroscopy data we performed low temperature specific heat measurements on the same $Fe(Sb_{1-x}Te_x)_2$ (\textit{x = 0.01, 0.05}) samples (Fig.4, upper panel). The data is in a rather good agreement with findings of Hu et al \cite{Hu_2012} on crystals from the same batch. Here power law divergences in $\gamma(T) = C/T$ and $\chi(T)$ are discussed in the framework of the disorder induced Griffith phase (GF) at the verge of magnetism \cite{Jaccarino_1967}. According to \cite{Castro_Neto_1998}, the low temperature divergence of specific heat in GF systems is described by power function $C(T)/T = a*T^{-1+\lambda_C}$ with $\lambda_C<1$. Then likewise \cite{Hu_2012}, the total low temperature behavior of specific heat in these compounds can be successfully fitted to the equation:
\begin{equation}
 C(T)/T = \alpha*T^{-1+\lambda_C}+b*T^2+c*T^4
\end{equation}
The second and third terms in Eq.2 describes harmonic and anharmonic contributions to specific heat, respectively \cite{0891168346}. The obtained values of $\lambda_C$ for both samples $(\lambda_C = 0.88$ (\textit{x = 0.01}) and \textit{0.70} (\textit{x = 0.05}) see tab 2) are in good agreement with that reported in Ref.\cite{Hu_2012}.
    \begin{table} [h]
        \caption{\label{tab:table 2}%
        Values of the parameter $\lambda$ obtained from specific heat ($\lambda _{C/T}$)$_{exp}$ and nuclear spin-lattice relaxation ($\lambda _{T1}$)$_{exp}$ experiments in comparison with that from Ref. \cite{Doniach1977}, marked by (*).
        }
        \begin{ruledtabular}
        \begin{tabular}{c|c|c|c}
        $x$&($\lambda _C $)$^*$&($\lambda _C$)$_{exp}$&($\lambda _{T_1}$)$_{exp}$\\ \hline
        0.01&0.91(7)&0.88(4)&0.90(6)\\ \hline
        0.05&0.72(3)&0.70(4)&0.64(4)\\
        \end{tabular}
        \end{ruledtabular}
    \end{table}

\section{Discussion}
The origin of the observed low frequency shoulder at $^{121,123}Sb$ NQR lines can be understood as follows. With increasing of $Te$ content \textit{x} from \textit{0.01} to \textit{0.05} the number of hetero-dumbbells $Sb$-$Te$ also increases. These dumbbells are characterized by polarization of the $Sb$-$Te$ bond due to higher electronegativity of the $Te$ atom. Therefor electronic density inside the $Sb-Te$ dumbbell is shifted towards $Te$ atom. As a consequence, a partial negative charge on $Sb$ atom is reduced causing decrease of EFG followed by decreasing of $Sb$ quadrupole frequency. At low $Te$ concentration (\textit{x = 0.01}) this effect in not yet visible but \textit{5\%} $Te$ seems to be enough for detection since the number of hetero-dumbbells $Sb$-$Te$ increases substantially and the left shoulder on $Sb$ NQR appears. In this simplified approach only the \textit{1-st} coordination sphere of $Sb$ is considered which might be visualized as a $Sb$-$Sb$ dumbbell with short interatomic distance ($\sim$ 2.8 \AA) surrounded by \textit{6} $Fe$ atoms likewise strongly distorted octahedron (Fig.5). In \textit{5\%} $Te$ substituted $FeSb_2$ sample one can expect appearance of hetero-dumbbell $Sb$-$Te$ in the \textit{2-nd} coordination sphere of homo-dumbbell $Sb$-$Sb$ which slightly reduce the charge on $Sb$. In conjunction with strong NQR line broadening caused by lattice disorder this explains why instead of separate peak to the left of main $Sb$ NQR line we observe just a left shoulder.
\begin{figure}[h]
\includegraphics[width=\linewidth]{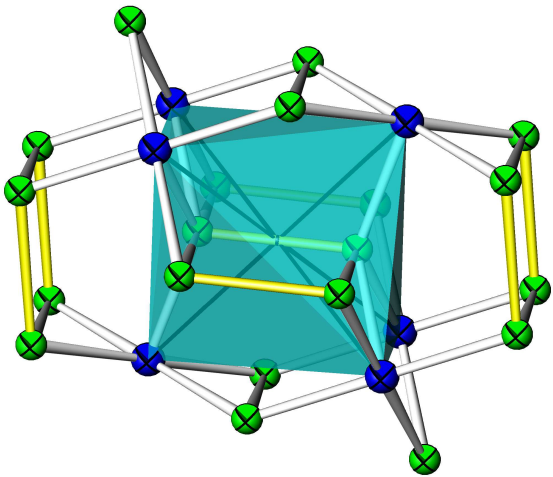} 
\caption{\label{fig:epsart} (color online). Schematic illustration of the first two coordination spheres of $Sb$-$Sb$ dumbbell in the $FeSb_2$ crystal structure. $Sb$ atoms are indicated as green balls, $Fe$ – dark blue balls. $Fe_6$ octahedron surrounding the center of $Sb$-$Sb$ dumbbell (black point in the center of octahedron) is indicated by green-blue.}
\end{figure}

    As seen from Table 1, for all samples the broadening of the $\nu_1$ line for $^{123}Sb$ isotope is higher than that for $^{121}Sb$ isotope in satisfactory accordance to the ratio of their quadrupole moments $^{123}Q/^{121}Q = 1.36$. This result provides an evidence of electronic quadrupole origin rather than magnetic origin of the $^{121,123}Sb$ NQR line broadening in $Fe(Sb_{1-x}Te_x)_2$ (\textit{x = 0.05}) sample. This supports the claim of electronic Griffith phase (EGP) in $Te$ doped $FeSb_2$ and is in a strong contrast to the magnetic Griffith phase (MGP). Indeed in case of isolated magnetically ordered clusters characteristic for the magnetic Griffiths phase $Sb$ nuclei inside these clusters should exhibit strong hyperfine magnetic fields of about $0.1 \div 1$ T induced from electron spins localized on $Fe$. Than instead of pure NQR one should observe Zeeman perturbed NQR on $^{121,123}Sb$ nuclei with pronounced splitting (or, at least, strong broadening) of initial NQR transition lines which depends on value and orientation of internal magnetic field in respect to the main EFG axes and asymmetry parameter $\eta$ and is proportional to gyromagnetic ratio $\gamma$ (see, for instance, \cite{Curro_2000}). Since $^{121}\gamma/^{123}\gamma = 1.85$ broadening of $^{121}Sb$ NQR lines should be almost twice as for $^{123}Sb$ isotope. This is definitely not seen in our experimental $^{121,123}Sb$ NQR spectra in $Fe(Sb_{0.95}Te_{0.05})_2$ sample. In the upper limit of hyperfine magnetic field ($\sim 1 $ T) induced on $Sb$ nuclei within ordered spin clusters of Griffiths phase one can even observe a "wipe-out" effect of disappearing of $Sb$ NQR lines originated from cluster volume due to their extreme Zeeman broadening. This effect should substantially reduce the total $Sb$ NQR intensity which was not observed in our experiment.
\newline \indent For a metal in the frame of the Landau Fermi liquid (LFl) the SLRR could be related to the specific heat via the density of states at the Fermi level which yields $1/T_1T \sim N^2(E_F) \sim \gamma^2 \equiv (C/T)^2 \sim T^{2(-1 + \lambda)}$. Contrasting to that, if the metal is a weak itinerant magnet, the SLRR is more related to the low energy and $q$- averaged complex dynamic susceptibility $\chi(q.\omega)$ which yields $1/T_1T \sim \sum\nolimits_{q} \chi (q. \omega ) $. Here it matters if the correlations are \textit{fm} (at \textit{q = 0)} of \textit{afm} (at $q\ \neq \ 0$). For \textit{fm} correlations the SLRR is frequently found to be proportional to the bulk susceptibility   $1/T_1T \sim \chi \sim T^{-1+\lambda}$ for the MGP.
    \newline \indent For the \textit{1\%} sample the specific heat coefficient power law ($m \equiv \ 1\ -\ \lambda \ =\ 0.12$) suggests a SLRR power law with \textit{n = 2m = 0.24} which is in rather good agreement with the experimental result (\textit{n = 0.2}). For the \textit{5\%} sample the specific heat coefficient power law (\textit{m = 0.28}) suggests a SLRR power law with \textit{n = 2m = 0.56} which is much smaller than what is found by experiment (\textit{n = 0.72}).  This might point towards the fact that upon doping we have a crossover from more localized correlated metal to and \textit{afm} correlated itinerant metal at the verge of order. Here Moryia predicted a power law with \textit{n = 3/4} which is rather close to the experimental finding.
\newline \indent Nonetheless the specific heat coefficient enhancement factor at 2 K  ($\gamma_{5\% }/ \gamma_{1\%} $) is about 5 which suggests in the LFl theory an enhancement of the SLRR (\textit{R = }$1/T_1T$) ($R_{5\%} / R_{1\%}$) $\approx$ 25 which is indeed experimentally confirmed by our spin lattice relaxation measurements.
    \newline \indent Let’s take a closer look to predictions for the SLRR in an Griffith phase. According to theoretical prediction for the magnetic Griffith phase the nuclear spin-lattice relaxation rate should follow the equation \cite{Castro_Neto_1998}:
    \begin{equation}
    1/T_1T(\omega , T) \varpropto \omega^{-2+\lambda} \tanh{(\omega/T)}
    \end{equation}

Since in our NQR experiment $\hbar \omega \ll k_B T$ even at lowest temperatures $\tanh(\omega/T)\approx\omega/T$ and Eq.(3) is simplified to the form:
\begin{equation}
1/T_1T(\omega , T) \varpropto \omega^{-1+\lambda}/T \varpropto T^{-1} 
\end{equation}

     The MGP power law with \textit{n = 1} (assuming $\omega$ = constant in the first place) is far from the experimental values (\textit{n = 0.2} (for \textit{x = 0.05}) and \textit{n = 0.72} (for \textit{x = 0.05})).

From the other hand, $\alpha \equiv 1 - n\ =\ 1/3$ is a characteristic exponent value within the Tsvelik and Reizer model based on scaling analysis of collective bosonic modes of the fluctuations with the spectrum $\omega\sim q^3$ near QCP providing $1/T_1T \sim T^{-2/3} \sim T^{-0.66}$ $(1/T_1 \sim T^{1/3})$ behavior at low temperatures \cite{Tsvelik_1993}.

     The SLR results obtained for the for $Fe(Sb_{1-x}Te_x)_2$ (\textit{x = 0.01}, \textit{0.05}) samples do not provide an unambiguous microscopic evidence in favor of either of these two models describing complicated NFL properties in the vicinity of the QCP. To gain an extra argument favoring one of these models we revisited the low temperature specific data analysis. Although approximation of specific heat experimental data by Eg. (2) is almost perfect above 5 K, the low temperature part of theoretical curve for the \textit{x = 0.05} sample shows systematic deviation from experimental $C(T)/T$ points which evidently diverges faster with decreasing T than power function $T^{-1+\lambda}$  (see Fig.4, lower panel). One should also take into account that more general description of NFL behavior predicts logarithmic rather than power divergence of specific heat at $T \rightarrow 0$ \cite{Stewart_2001}. It is worth to note that the Tsvelik and Reizer model also predicts for the specific heat low temperature logarithmic divergence \cite{Tsvelik_1993} However, $-\ln{T}$ function diverges even slower than $T^{-1+\lambda}$. This contradiction can be settled by implementing a dissipative quantum droplet model \cite{Castro_Neto_2000} which describes the critical behavior of NFL metallic magnetic systems at low $T < T^{*}$ below which quantum critical regime is dominated by dissipation providing stronger divergence of specific heat than the power law: $C(T)/T \varpropto T/ln^2{(1/T)} $. Combining general logarithmic NFL divergence with even stronger divergence term of $C(T)/T \varpropto T/ln^2{(1/T)} $ from dissipative quantum droplet model one arrives at:
    \begin{equation}
    C(T)/T=a*/T\ln^2{(1/T)}-b*\ln{T}+c*T^2+d*T^4
    \end{equation}

Using Eq.5 instead of Eq.2 we obtained much better agreement with experimental low temperature $C$($T$)/$T$ data as demonstrated in Fig.4, lower panel.

\section{Summary}
We performed a comprehensive study of correlated intermetallic system $Fe(Sb_{1-x}Te_x)_2$ (\textit{x=0.01, 0.05}) in the vicinity of an antiferromagnetic quantum critical point by means of NQR spectroscopy on $^{121,123}Sb$ nuclei. It was found that even a slight Tellurium doping of \textit{x=0.05} introduces strong lattice disorder in the binary Kondo-insulator compound $FeSb_2$ resulting in substantial asymmetric broadening $Sb$ NQR spectrum and formation of the low frequency shoulder at left side of each of $^{121,123}Sb$ NQR transition lines due to polarization of the $Sb$-$Te$ bond and shifting of electronic density inside the $Sb$-$Te$ dumbbell towards $Te$ atom. Furthermore the observed transformation of the $Sb$ NQR spectrum in $Fe(Sb_{1-x}Te_x)_2$ samples are related to local changes of the electric field gradient due to the doping effect. There is no evidence for magnetic broadening of the NQR lines due to the emerging $Fe\ -$ magnetism upon doping. We interprete this as the microscopic evidence for the electronic Griffith phase in strong contrast to the magnetic broadening expected for the magnetic Griffith phase.

    The spin lattice relaxation results clearly show that the charge gap of the pure correlated semimetal $FeSb_2$ is filled upon $Te$ doping. In a first approximation based on the Landau Fermi liquid theory for correlated metals the low temperature divergence of the SLRR $1/T_1T(T)$ could be scaled to the one of the specific heat coefficient $\gamma (T)$. A very good agreement was found for the \textit{1\%} sample wheras the \textit{5\%} sample which is closer to the antiferromagnetic ordered phase shows a significant violation of the scaling. The power law coefficient \textit{n = 0.72} is rather close to the one expected for antiferromagnetic criticality which is $n_{afm}\ =\ 3/4\ =\ 0.75$. Nonetheless the enhancement factor of \textit{5 } between the \textit{x = 0.01}  and \textit{x = 0.05} in the specific heat yields an enhancement of \textit{25} in the $1/T_1T$ value which was experimentally verified. As for both samples the the specific heat divergence is in good agreement with Ref \cite{Hu_2012} which suggests a electronic Griffith phase the microscopic SLRR shows a remarkable lack of consistency with the magnetic Griffith phase predictions. Probably this is because of the $q-$ averaging nature of the SLRR. Nonetheless strong evidence for antiferromagnetic critical fluctuations at zero field for the \textit{5\%} $Te$ doped $FeSb_2$ sample is given. Antiferromagnetic criticality is a rare occurrence in $Fe-$based systems and therefore doped $Fe-$based semimetals in general might provide a platform for further studies.
Further more other local probes (like $\mu $SR) should be addressed to study Griffith phase systems.

\section{Acknowledgments}
Authors are grateful to A. V. Shevelkov and E. A. Kravchenko for fruitful discussions, and A. V. Trofimenko for help with the experiment. A. A. Gippius and S. V. Zhurenko appreciate financial support from Russian Foundation of Basic Research (grant 16-53-52012-a). We thank H. Rave and C. Klausnitzer for technical support. Work at BNL was supported by the U.S. DOE-BES, Division of Materials Science and Engineering, under Contract No.DE-SC0012704.


\bibliography{Paper_FeSb2Te_V5_2}

\end{document}